# Vanishing biexciton binding energy from stacked, MOVPE grown, site-controlled pyramidal quantum dots for twin photon generation

S. T. Moroni, S. Varo, G. Juska, T. H. Chung, A. Gocalinska, E. Pelucchi

**Abstract**

We characterized stacked double-pyramidal quantum dots which showed biexciton binding energies close to zero by means of photoluminescence and cross-correlation measurements. It was possible to obtain a sequence of two photons with (nearly) the same energy from the biexciton-exciton-ground state cascade, as corroborated by a basic rate-equation model. This type of two-photon emission is both of relevance for fundamental quantum information theory studies as well as for more exotic applicative fields such as quantum biology.

**1. Introduction**

Semiconductor quantum dot (QD)-based light sources are capable of delivering quantum light of various nature and are intensively studied for both fundamental and technological purposes. For instance, state of the art single QDs are being developed pursuing pure single photon emission[1][2][3], generation of highly indistinguishable photons[4][5], production of entangled photon pairs with high fidelity[6][7][8], and also exploitation of time-bin entanglement ([9] and references therein). All these are relevant for applications in quantum information and computation implementations based on photonics [10][11], or with a strong photonics presence. Among the spectrum of conceivable non-classical light, "twin" photons (i.e. identical photons pairs, with the same energy) represent an interesting possibility. Applications have been proposed in the field of quantum information itself as well as in other more "exotic" areas, such as quantum biology[12].

It is well known that this type of two-photon state can be generated by means of spontaneous parametric down conversion, relying on probabilistic emission rather than on a truly quantum mechanical "on-demand" scheme. Moreover, only a few cases are reported in the literature, where twin photon generation was achieved on an "integrated" platform. It is in fact possible to produce twin photons from a semiconductor parametric-down-conversion-based integrated system[13], although with strong compromises on efficiency. More recently, III-V semiconductor QDs were employed to this purpose, where the two photons generated through the biexciton-exciton-ground state cascade can have the same energy when the biexciton binding energy is zero[14]. As we will show in the following, generation of two photons with the same energy is possible by the excitation of pyramidal quantum dot (PQD) stacked systems by employing alternative epitaxial growth strategies from those already proposed, despite benefitting from remaining in the same III-V family which has already delivered high quality results when quantum-light is to be considered ([1][6][8]).

Engineering the surface of a growth substrate for the fabrication of QDs can be extremely beneficial to develop technological alternatives and to advance quantum optics/technologies. For instance, although state of the art self-assembled QDs were widely employed to reach milestones in this field (as per the above mentioned references), control over the position together with the overall structural properties of the emitters can only be achieved using selective-area growth and/or other pre-patterning strategies. All, not necessarily in the same material system obviously (see e.g.[15] [16][17] which could also potentially obtain similar results), including exploring dots in nanowires for example (e.g. [18][19]), i.e. a rather different approach from conventional self assembled dots. In this context the PQD system is an outstanding example of how metalorganic vapour-phase epitaxy (MOVPE) can be employed to reach high control over the structural and optical properties of QDs, and how this has been indeed possible thanks to deep understanding of the growth processes [20][21]. PQDs have been proved to allow high density of entangled photon emitters by both optical [22] and electrical excitation [23]. But one of the unique capabilities offered by the PQD system is that of systematically stacking two or more QDs on the top of each other, maintaining near identical structures, and allowing for the tuning of their optical properties, among which the biexciton binding energy, as we have recently showed [24].

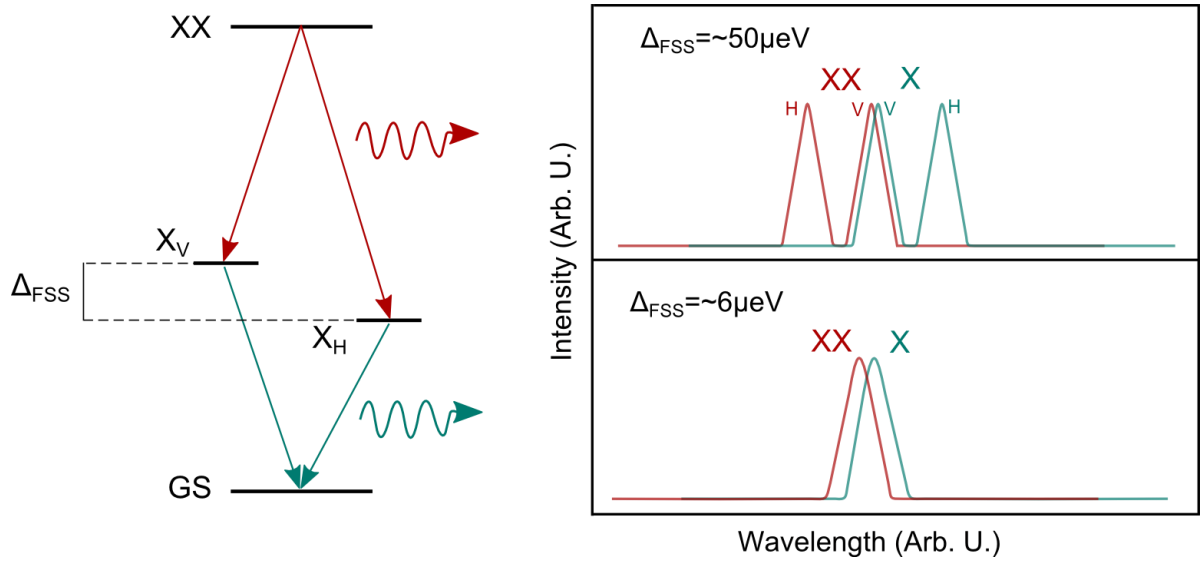

Fig.1 – On the left: general schematic representation of Fine-Structure-Splitting (FSS) in the biexciton-exciton-ground state of a QD; on the right: schematic representation of the spectra of biexciton (red) and exciton (green) transitions for QDs with relatively small biexciton binding energy and significant FSS (top) as in [14], and for QDs with vanishing biexciton binding energy and small FSS (bottom) as in our stacked PQDs.

Compared to the only other case of QD-based twin photon generation we are aware of presented in the literature[14], PQDs present two advantages: the first is technological, as PQDs allow for precise positioning of the source, an important requirement for practical implementations. The second bears a more fundamental aspect. In fact, in ref.[14] the authors exploited the natural asymmetry of their QD system, delivering a split in the exciton and biexciton states (normally referred to as fine structure splitting, see Fig 1 where Ref. [14] and our approach are compared) to cancel out the relative binding energies. Therefore they selected only one polarization, filtering 50% of the events of photon generation. This is not (at least ideally) the case we're presenting, where all the photons are potentially contributing.

## 2. Experimental

PQDs are fabricated starting from a patterned (111)B-oriented GaAs substrate (in which tetragonal recesses are obtained), on which MOVPE is performed depositing InGaAs dots with GaAs barriers, as for example discussed in ref [25] and [26]. The growth process is the interesting result of the competition between precursors decomposition rate anisotropies, and adatom diffusion and preferential attachment at concave recesses (also referred to in the literature as capillarity processes) [27][28]. More details on the fabrication procedure of the stacked system can also be found in ref. [24]. Samples also undergo a "back-etching" process to turn the pyramids pointing up by removing the original growth substrate, allowing an enhanced light extraction. The optical characterization of the samples presented in this paper was performed by photoluminescence and photon correlation spectroscopy using a standard HBT setup. All measurements were performed at 10K in a non-resonant excitation scheme, as described elsewhere [22].

## 3. Results and discussion

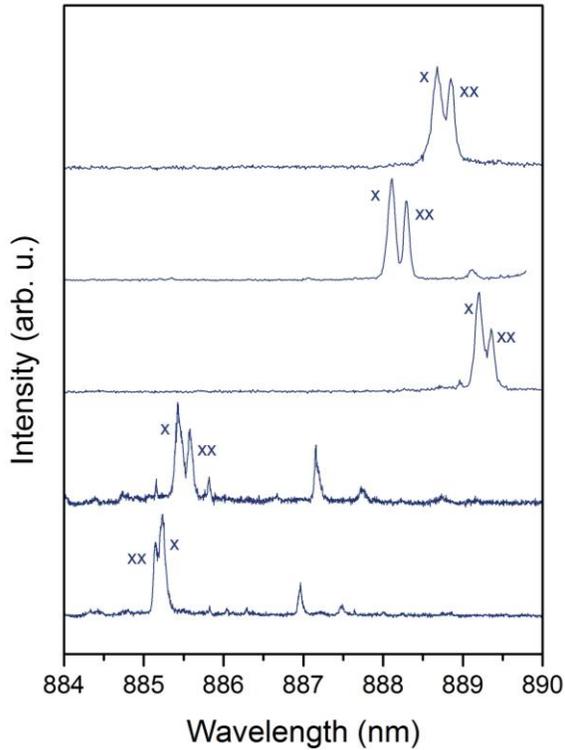

Fig.2 – A collection of spectra from the same sample with 2 QDs with 1nm inter-dot barrier showing little or vanishing biexciton binding energy. Exciton and biexciton are marked with the labels "X" and "XX" respectively in each spectrum.

We previously reported on the effect of stacking PQDs in the same pyramidal structure [24]: as the distance between nominally identical QDs (i.e. with the same thickness) was reduced, a previously unreported regime was entered of only single dot "like" emission. Also a red-shift of the emission was systematically reported as a function of reducing inter-dot layer thickness, together with an overall change in the binding energy of the biexciton (normally, and here, defined as the energy difference between exciton and biexciton transitions). For instance it was possible to gradually go from samples with anti-bonding biexciton (binding energy<0) to samples with bonding biexciton (binding energy>0) by simply varying the distance between the dots. We pause here a second, to stress the relevance and unique control demonstrated by these results. Indeed the peculiarity of the MOVPE III-V growth process utilized here [20][29] allows what is a uniquely demonstrated control over uniformity and dot reproducibility [30], which should be ascribed to the distinctive MOVPE processes involved in a concave environment (i.e. decomposition rate anisotropy and specific adatom diffusion/capillarity processes/anisotropies).

The sample with inter-dot barrier of 1 nm resulted in a distribution of binding energies with an average of 0.2823 ± 0.43254 meV, interestingly allowing to easily find quantum dots with nearly zero biexciton binding energy by "scanning" the QD ensemble/sample.

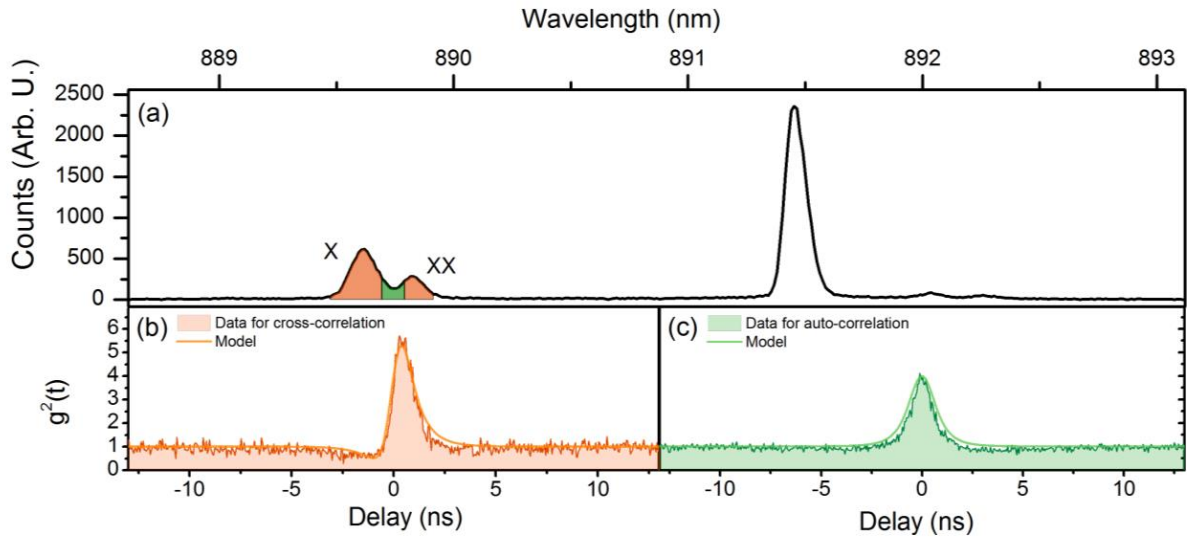

Fig.3 – a) A representative spectrum for a typical PQD with 2 QDs at 1nm inter-dot barrier; exciton and biexciton are labeled with X and XX respectively (while the brighter peak at longer wavelengths is relative to a negatively charged exciton[25]) Auto-correlation for the photons in the spectral overlap between the two excitonic transitions (schematically highlighted with the green color in a)) showing a significant bunching and a symmetric shape, as found in similar cases in literature for twin photons [14]. c) Cross-correlation between biexciton (start channel of the correlator) and exciton (stop channel), highlighted with the same color in a), showing strong bunching, confirming the cascade between the two transitions.

We concentrated our characterization on the QDs with nearly-zero biexciton binding energy from the 1 nm inter-dot barrier sample. Although a complete overlap between exciton and biexciton transitions was statistically hard to find (and eventually even harder to recognize as two superimposed transitions by simply observing the spectrum), it was possible to find PQDs where significant overlap between the two excitonic transitions could be observed. We note also that, on single PQDs, a similar zero binding energy was also reported in the past, but no experimental optical correlation verification actually performed [31]. Also it should be said that spectral wandering resulted in average linewidths of exciton and biexciton of respectively 172μeV and 119μeV. Fig.2 shows a collection of spectra from this sample where the biexciton binding energy was small and exciton and biexciton were showing a significant spectral overlap. We stress that it is easy to locate quantum dots showing this characteristics on the sample under analysis. Moreover, we need to specify at this point that the large overlap between the exciton and biexciton spectra is also due to the relatively large linewidths of the peaks. The origin of the spectral wandering is to be found in the non-resonant excitation technique employed, causing therefore the small FSS to be hidden. In case of resonant or quasi-resonant excitation of the QD we would expect the linewidth to sensibly decrease and partially reduce the overlap between the transitions, also revealing the small FSS. We note that this can/could be corrected in a second step by employing tuning strategies, such as the application of piezoelectric stress [32].

We consider here as representative/significant example the spectrum of a QD presented in Fig.3a: in this case it was possible to spectrally separate photons from each transition with our spectrometer while still having a significant overlap between the exciton and biexciton (which were identified using power dependence and cross-correlation, following criteria described in [24]). Fig.3b shows the cross-correlation experiment between biexciton and exciton, performed by selecting the "outer" part of each peak with respect to the overlap, having set-up a resolution of our spectrometer of about 0.2nm. This resulted in a typical bunching, with a $g^2(t)$ up to 6, also demonstrating the biexciton-exciton cascade.

We also performed autocorrelations energetically filtering only photons relative to the spectral overlap of the two transitions; Fig.3c shows the autocorrelation measurement for the selected QD. A symmetric $g^2(t)$ was collected, showing a pronounced bunching at zero delay time and a slight anti-bunching at a few nanoseconds delay. We recognize this as a signature of the emission of two subsequent photons in the selected energy window, as reported previously in the literature [14]. Indeed, the total $g^2(t)$ is given by the sum between the four possible cross-correlation collection events (exciton autocorrelation, biexciton autocorrelation, and the two cross-correlations between exciton and biexciton with reversed order), when either exciton or biexciton are collected at each of the two photodetectors.

We employed a simple rate-equation model to reconstruct the total $g^2(t)$ starting from the measured lifetimes for exciton and biexciton transition: by approximating the system to a 3 level cascade (biexciton-exciton-ground state) for simplicity, we computed each of the $g^2(t)$ functions for exciton autocorrelation, biexciton autocorrelation, exciton and biexciton cross-correlation (in both orders), also taking into consideration the characteristic response of the measurement apparatus. The only free parameter being the capture rate for an electron-hole pair, we fitted the cross-correlation data and plotted the total $g^2(t)$ autocorrelation for the photons coming from the overlap. Fig.2 shows the fitted data (continuous curve) overlapping the data with good approximation. Although the calculated values for the $g^2(0)$ are in good agreement with the experiment, the full simulated curve slightly deviates from the experiment, suggesting that part of the recombination dynamics (e.g. charged complexes, low-intensity slow lifetimes components) would have to be included in a more complete model. We also note that in order to achieve this fit it was necessary to assume a measured "bad" single-photon emission from both exciton and biexciton, expressed as a "bad" $g^2(0)$ in the corresponding autocorrelations (bad meaning an autocorrelation $g^2(0)$ bigger than 0.5), as typically observed on this sample, probably due to fast re-excitation events.

While in this experiment only part of the photons emitted through the biexciton cascade are falling in the overlapping spectral window, our result act as conceptual proof-of-concept, especially in view of the fact that it should be possible, in the future, to use methods that allow a small tuning of the binding energy, in order to achieve a perfect overlap between the transitions for each of the quantum dots. For example in this contest, the application of a stress to the whole pyramidal structure by means of piezoelectric devices[33].

In conclusion, we showed that MOVPE growth processes combined with patterning of the substrate allow for the seeding and site-control of PQDs with a double-QD structure, presenting single dot like spectra and also allowing the formation of biexcitons with quasi-zero binding energy when proper dot dimensions are carefully chosen/engineered. Photon statistics for the photons coming from the spectral overlap between exciton and biexciton corresponds to that expected for two consecutive photons with the same energy. The reduction of the residual binding energy by means of tuning strategies combined with resonant excitation techniques would easily allow, for example, to have a larger amount of photons with the same energy, while resonant pumping techniques should also allow for indistinguishable photons[8] and an "ideal" twin-photons generation.

**Acknowledgments**

This research was supported by Science Foundation Ireland under Grant Nos. 10/IN.1/I3000, 15/IA/2864, and 12/RC/2276. The authors are grateful to Dr. K. Thomas for the MOVPE system support.

**Rate equation modeling (supplementary material)**

We employed a three-level rate equation model to describe the temporal evolution of the excitonic complexes in the PQD. The set of equations can be written as:

$$\frac{d\vec{p}(t)}{dt} = \begin{bmatrix} -1/t_{eh} & 1/\tau_X & 0 \\ 1/t_{eh} & -1/t_{eh} - 1/\tau_X & 1/\tau_{XX} \\ 0 & 1/t_{eh} & -1/\tau_{XX} \end{bmatrix} \begin{pmatrix} p_0(t) \\ p_X(t) \\ p_{XX}(t) \end{pmatrix}$$

where $\vec{p}(t) = (p_0(t), \; p_X(t), \; p_{XX}(t))$ is the time-dependent occupation probability for each level (biexciton (XX), exciton (X) and ground state (0)), $t_{eh}$ is the capture time for one electron-hole pair, $\tau_X$ and $\tau_{XX}$ are exciton and biexciton lifetimes respectively. Once the initial conditions are set for a transition to happen at time $t = 0$, one can compute the autocorrelation function $g^2(t)$ for that specific transition [34]. For exciton decay we have the initial condition $\vec{p}(0) = (1,0,0)$, for biexciton decay $\vec{p}(0) = (0,1,0)$. The $g_i^2(t)$ for the i-th transition is expressed as $g_i^2(t) = p_i(|t|)/p_i(\infty)$. Cross-correlation is given by $g_{X-XX}^2(t) = p_X(t)/p_X(\infty)$ and $\vec{p}(0) = (0,1,0)$ for t>0, $g_{X-XX}^2(t) = p_{XX}(-t)/p_{XX}(\infty)$ and $\vec{p}(0) = (1,0,0)$ for t<0. The solution for each correlation function is convoluted with the total response function of the detection setup (a Gaussian with 400ps FWHM).

In our simulation $\tau_X$ and $\tau_{XX}$ were fixed, their values estimated from direct measurement, while $t_{eh}$ was the only free parameter, which was varied to fit the cross-correlation function $g_{X-XX}^2(t)$. Having these parameters, all the possible $g^2$ functions could be plotted. A linear combination of these resulted in the graph of Fig.3c.